\documentclass[a4paper,10pt,onecolumn,preprint]{elsarticle}
\usepackage[latin2]{inputenc}
\usepackage[mathscr]{eucal}
\usepackage{amssymb,amsthm,amsmath,enumerate}
\usepackage{graphicx, graphics}
\usepackage{ifthen}
\usepackage{array}
\begin{document}

\theoremstyle{plain}
\newtheorem{theorem}{Theorem}
\newtheorem{lemma}{Lemma}
\newtheorem{proposition}{Proposition}
\newtheorem{corollary}{Corollary}
\theoremstyle{remark}
\newtheorem{definition}{Definition}
\newtheorem{remark}{Remark}
\newcommand{\sgn}{{\mathrm\,sgn\,}}
\newcommand{\diag}{{\mathrm\,diag\,}}
\newcommand{\bR}{{\mathbb R}}
\newcommand{\arccot}{\operatorname{arccot}}

{\begin{frontmatter}

     \title{Dynamical properties of a nonlinear growth equation}

\author[UPJV1]{Mohammed Benlahsen}
\author[ME]{Gabriella Bogn\'ar \corref{cor1}}
\ead{matvbg@uni-miskolc.hu}
\author[UPJV2]{Mohammed Guedda}
\author[ME]{Zolt\'an Cs\'ati}
\author[ME]{Kriszti\'an Hricz\'o}


\cortext[cor1]{Corresponding author}

\cortext[cor2]{Principal corresponding author}

\fntext[fn1]{This is the specimen author footnote.}
\fntext[fn2]{Another author footnote, but a little more longer.}
\fntext[fn3]{Yet another author footnote. Indeed, you can have any
number of author footnotes.}

\address[UPJV1]{LPMC, Universit\'e de Picardie Jules Verne, Amiens, France}
\address[UPJV2]{LAMFA, UMR-CNRS 7352, Universit\'e de Picardie Jules Verne, Amiens, France}
\address[ME]{University of Miskolc, Miskolc-Egyetemv\'aros 3515, Hungary}


     \begin{abstract}
       The conserved Kuramoto-Sivashinsky equation is considered as the
evolution equation of amorphous thin film growth in one- and in
two-dimensions. The role of the nonlinear term $\Delta (\left\vert
\nabla u\right\vert ^{2})$ and the properties of the solutions are
investigated analytically and numerically. We provide analytical
results on the wavelength and amplitude. We present numerical
simulations to this equation which show the roughening and
coarsening of the surface pattern and the evolution of the surface
morphology in time for different parameter values in one- and in
two-dimensions.
     \end{abstract}


  \end{frontmatter}
}




\date{\today}

\section{Introduction}
Molecular Beam Epitaxy (MBE), which has many important technological
and industrial applications, is often used to grow nanostructure on
crystal surfaces. The evolution of the surface morphology during MBE
growth results from a competition between the molecular flux and the
relaxation of the surface profile through surface diffusion of
adatoms. One crucial aspect of the growth process is its possible
unstable character,  due to deterministic mechanisms,  which prevent
the growing surface to stay parallel to the substrate
\cite{V1}-\cite{ETB}.

 This phenomenon has turned out to be a source of a wide class of  nonlinear dynamics, which varies from
spatio-temporal chaos \cite{BMV} to the formation of stable
structures \cite{US}, from coarsening processes \cite{PGPM} to
diverging amplitude structures \cite{PM}. One of the many challenges
involved in applied mathematics and nonequilibrium physics is to
predict the behavior of surface evolution, from the knowledge of an
initial arbitrary profile, and the scaling relationships between
surface features in various growth regimes. In \cite{[6]}, Frisch
and Verga studied the step meandering instability on a surface
characterized by the alteration of terraces with different
properties as in the case of Si(001). Under the assumption of
negligible desorption and Erlich-Schwoebel (ES) effect, the surface
morphology is investigated by means of the following unstable mode
equation (Conserved Kuramoto-Sivashinsky, CKS for short)
                \begin{equation}
                \partial_tu = -\partial_x^2\left[\nu u+\kappa\partial_x^2u+\mu \left(\partial_x u\right)^2\right]. \label{eq1}
                \end{equation}
The unknown function $u(x,t)$ designates the amplitude of the
unstable branch, $t$  is the time and $x$ is the coordinate along
the step. Coefficients $\nu, \mu $ and $\kappa\  (\kappa = 1$) are
positive physical  parameters.

Equations of the type (\ref{eq1}) have been employed in different physical contexts.  In Ref.~\cite{[10]}, the authors mentioned that equation (\ref{eq1}) is a possible (natural) candidate for the time-evolution of the meandering amplitude, if desorption is negligible.\\
The CKS equation, after rescaling, can be interpreted as a
particular case  ($\epsilon = 0$) of the following modified CKS
equation
                \begin{equation}\partial_tu=-\partial_x^2\left[u-\epsilon \partial_x u+\partial_x^2u+\frac{1}{2}\left(\partial_x u\right)^2\right]. \label{eq3}
                \end{equation}
The above equation is used to describe bunches created by an
electromigration current~\cite{[5]}.
It is worth noticing that the term $\partial_x^3u$ can be removed
from equation (\ref{eq3}) via transformation
$u\rightarrow u-\epsilon x$. Equation (\ref{eq3}) is also proposed in~\cite{[11]} to describe sand ripples formation close to the instability threshold.\\

In 2009, Politi and ben-Avraham \cite{PobA} showed that the CKS
equation can be mapped into the motion of a system of particles with
attractive interactions, decaying as the inverse of their distance.

The following continuum $d-$dimensional model ($d=1,2$)
\begin{equation}\label{amorph}
\partial_t u =-\Delta\left[\nu u + \kappa\Delta u+ \mu \vert \nabla u\vert^2\right]
\end{equation}
with positive coefficients $\nu, \kappa $ and $\mu , $ has been
introduced by  Raible et al. \cite{RLH}  in the context of amorphous
thin film growth. The above equation  is a closely related to a more
general equation
\begin{equation}\label{sputt}
\partial_t u =-\Delta\left[\nu u + \kappa\Delta u+ \mu \vert \nabla u\vert^2\right]+\Lambda\vert \nabla u\vert^2,
\end{equation}
or to the following equation
\begin{equation}\label{sputt-scaling}
\partial_t u =-\Delta\left[u + \Delta u+ r\vert \nabla u\vert^2\right]+\vert \nabla u\vert^2.\end{equation}
Equation (\ref{sputt-scaling}), which  is deduced from (\ref{sputt})
with  $r = \nu \mu /\kappa \Lambda ,$ appears in the context of ion
beam sputtering (IBS). This equation is  obtained by Castro et al.
\cite{CCVG} from  a two-dimensional (reaction and transport
mechanisms) system of the coupled thickness of the mobility surfaces
adatoms layer and the height of the bombarded surface $u$ (see also
\cite{MCC} for one-dimension case). Note  that  for $r=0$ equation
(\ref{sputt-scaling}) reduces to the famous Kuramoto--Sivashinsky
equation which is known to produce spatio-temporal chaos.  For $r\to
\infty$ ($\Lambda\to 0 $), we obtain equation (\ref{amorph}) from
(\ref{sputt}). Therefore, one finds that $r$ is a very important
parameter, which determines the character of the solutions to
equation (\ref{sputt-scaling}).

Note that if $\kappa=\mu = 0$,
equation (\ref{sputt}) reads 
 \begin{equation}\label{G-KPZequation}
\partial_t u = -\nu\Delta u + \Lambda\vert\nabla u\vert^2,
\end{equation}
and then the new function $v\equiv -\frac{\Lambda}{\nu} u(x,-t/\nu)$
satisfies the well-known Kardar-Parisi-Zhang (KPZ) equation
\cite{TLDS}
 \begin{equation}\label{KPZequation}
\partial_t v = \Delta v + \vert\nabla v\vert^2,\end{equation}
which has explicit solution \cite{Gilding} given by
\begin{equation}\label{v}
v(x,t)=ln\left( \frac{1}{4\pi t} exp\left( -\frac{x^2+y^2}{4t}\right
) \right).
\end{equation}
If $\nu=\Lambda=0$ in (\ref{sputt}), we obtain, as above,  the
conserved Kardar-Parisi-Zhang (CKPZ) equation
\begin{equation}\label{KPZequation}
\partial_t v = \Delta\left(\Delta v + \vert\nabla v\vert^2\right).
\end{equation}

Results,  for the coarsening process, have been  presented for
(\ref{amorph}) and (\ref{sputt-scaling}) (see below), but do not
seem to describe completely the dynamics. The aim of this work is to
revisit, from the theoretical point of view,  equations
(\ref{amorph}) and (\ref{sputt-scaling}). In particular, we shall
present results  showing that surfaces can be mathematically and
physically  classified into different categories. Attention will be
focused on the effect of the CKPZ term $ \Delta(\vert\nabla
u\vert^2) $  for equation (\ref{amorph}) and  the interplay between
the CKPZ and the KPZ, $\vert\nabla u\vert^2,$ terms for equation
(\ref{sputt-scaling}).

\section{Analytical results}

In this section we analyze equation (\ref{amorph}) and for sake of
comparison we consider equation (\ref{sputt-scaling}) without the
KPZ term:
\begin{equation}\label{sputt-scaling-0}
\partial_t u =-\Delta\left[u + \Delta u+ r\vert \nabla u\vert^2\right].\end{equation}

For reader convenience, we analyze, as in \cite{RLH}, one and two
dimensional cases separately applying similarity method. Then, the
effect of the KPZ term is investigated.

\subsection{One-dimensional problem}

Here, we investigate the solutions of the conserved
Kuramoto-Sivashinsky (CKS) equation in one dimension.
\begin{equation}
u_{t}+\frac{\partial }{\partial x^{2}}\left[ u+u_{xx}+r \left(
u_{x}\right) ^{2}\right] =0.  \tag{CKS-1}  \label{CKS-1}
\end{equation}
As mentioned before the above equation was considered in \cite{[6]}.
Numerical solutions reveal that the typical length scale grows as
$\lambda(t)\sim t^\beta,$ with the coarsening exponent $\beta = 1/2$
(see also ~\cite{[5]}). It is shown  that  the general asymptotic
solution can be thought of as a superposition of parabolas.
Similarity solutions are also considered. Trying the solution ansatz
\begin{equation}
u(t,x)=t^{\alpha }f\left(\eta\right),\quad \eta= xt^{-\beta},
\label{1}
\end{equation}
one obtains
 $\alpha =1,$
$\beta =1/2$ and  we get
\begin{equation}
f-\frac{1}{2}\eta f^{\prime }+f^{\prime \prime }+t^{-1}f^{\prime
\prime \prime \prime }+r \left( f^{\prime 2}\right) ^{\prime \prime
}=0,  \label{10}
\end{equation}
where $^\prime=df/d\eta.$

Note that at $t=0$ we cannot eliminate $u_{xxxx}$. In general, there
is no similarity solution.

If $t\rightarrow \infty $ then we suppose that  $t^{-1}f^{\prime
\prime \prime \prime }\rightarrow 0$. Particularly, if
$f^{(4)}\equiv 0$, then $f=a\eta ^{3}+b\eta ^{2}+c\eta +e.$ From
equation (\ref{10}) the coefficients in $f$ are $a=0,c=0,e=e(b)$,
with some $b<0, e<0$
\begin{equation}
\begin{split}
u(x,t)=t\left( b\frac{x^{2}}{t}-e\right) =bx^{2}-et \\ =  -e\left(
t-\frac{b}{e} x^{2}\right)_{+}. \label{uxt}
\end{split}
\end{equation}

From (\ref{uxt}) one immediately sees that $\lambda
(t)=2\sqrt{\frac{e}{b}t}$ and $A(t)=\vert {e}\vert t$, i.e., $\alpha
=1$ and $\beta =1/2$.\\
We note that in \cite{[6]}, the authors obtained an explicit
solution having the form $u_{\rm ex}(x,t)=-x^2/(r4),$ for $\vert
x\vert < const.\sqrt{t},$ and zero elsewhere. Explicit solution
(\ref{uxt}) and $u_{\rm ex}$ indicate that for any initial
conditions $A(0)$ and $\lambda(0)$, the surface will grow without
limit (uninterrupted coarsening), which is physically not correct.

\begin{figure}
\begin{center}
    \includegraphics[keepaspectratio, width=4.5cm]{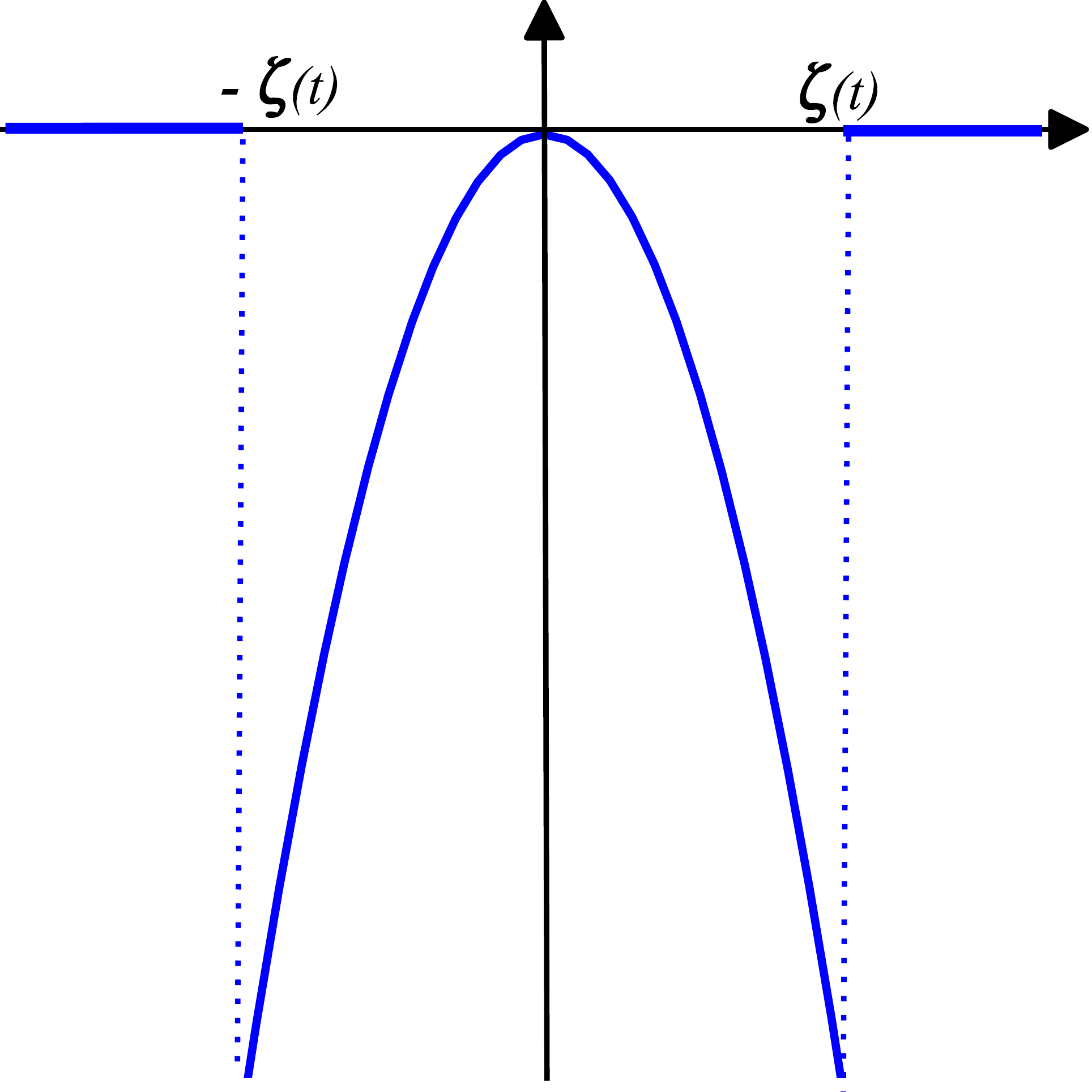}
    \caption{The height profile }
    \label{fig:1D_1aabra}
\end{center}
\end{figure}

Next, multiplying equation
\begin{equation*}
u_{t}=-u_{xx}-u_{xxxx}-r \left( \left( u_{x}\right) ^{2}\right)
_{xx}
\end{equation*}
by $u$ and integrating twice one gets
\begin{equation}
\frac{d}{dt}\frac{1}{2}\underset{\mathbf{R}}{\int }u^{2}dx=\underset{\mathbf{R}%
}{\int }\left( u_{x}\right) ^{2}dx-\underset{\mathbf{R}}{\int
}\left( u_{xx}\right) ^{2}dx.  \label{11}
\end{equation}
Shifting the maximum of the unsteady solution to $x=0$ from equation
(\ref{CKS-1})  we get the form of $u$ as
\[
u(x,t)=\left\{
\begin{array}{c}
-\frac{1}{4 r }x^{2}\text{ if }\left\vert x\right\vert <\zeta (t), \\
0\text{ \ \ \ if \ \ \ \ }\left\vert x\right\vert \geq \zeta(t),%
\end{array}%
\right.
\]%
where $\zeta (t)$ is an unknown function (see
Fig.\ref{fig:1D_1aabra}).

\begin{figure}
\begin{center}
    \includegraphics[keepaspectratio, width=7cm]{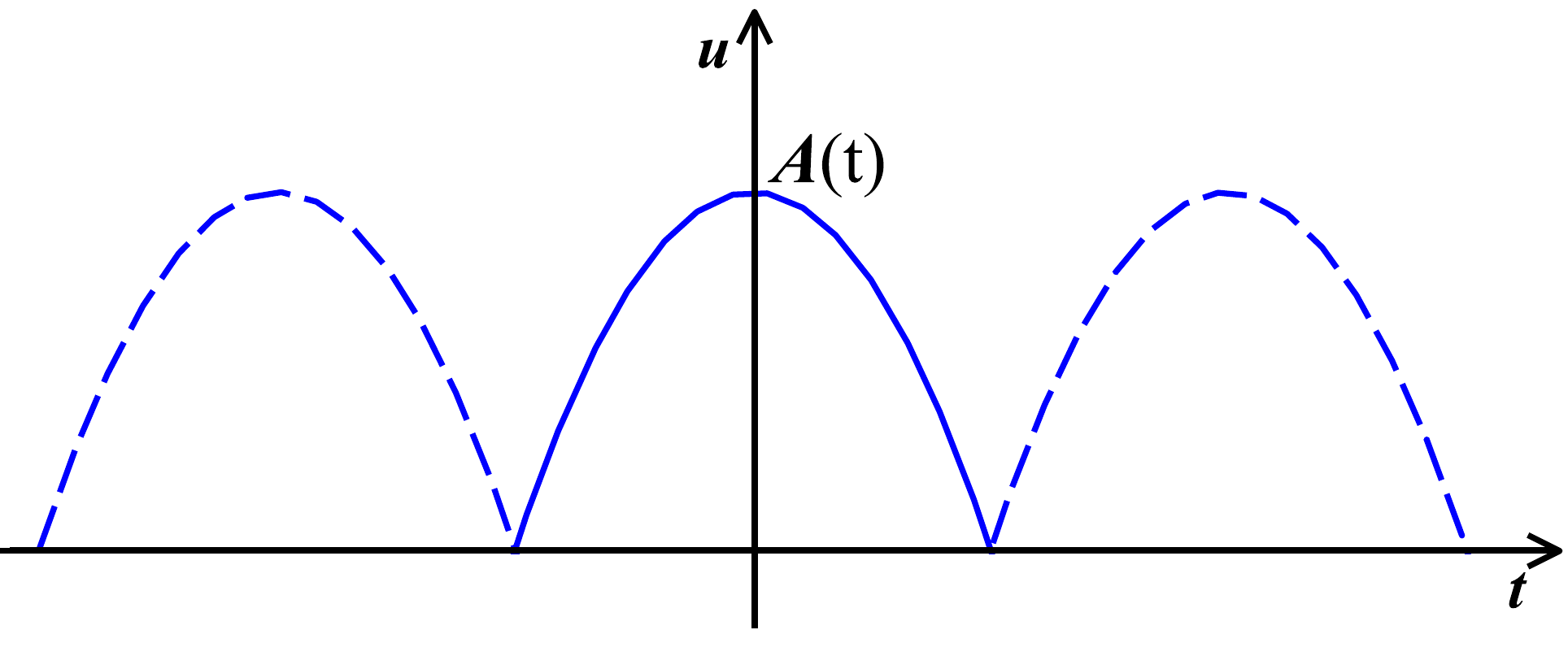}
    \caption{The height profile of $u(x,t)$}
    \label{fig:1D_1abra}
\end{center}
\end{figure}

\begin{figure}
\begin{center}
    \includegraphics[keepaspectratio, width=4.5cm]{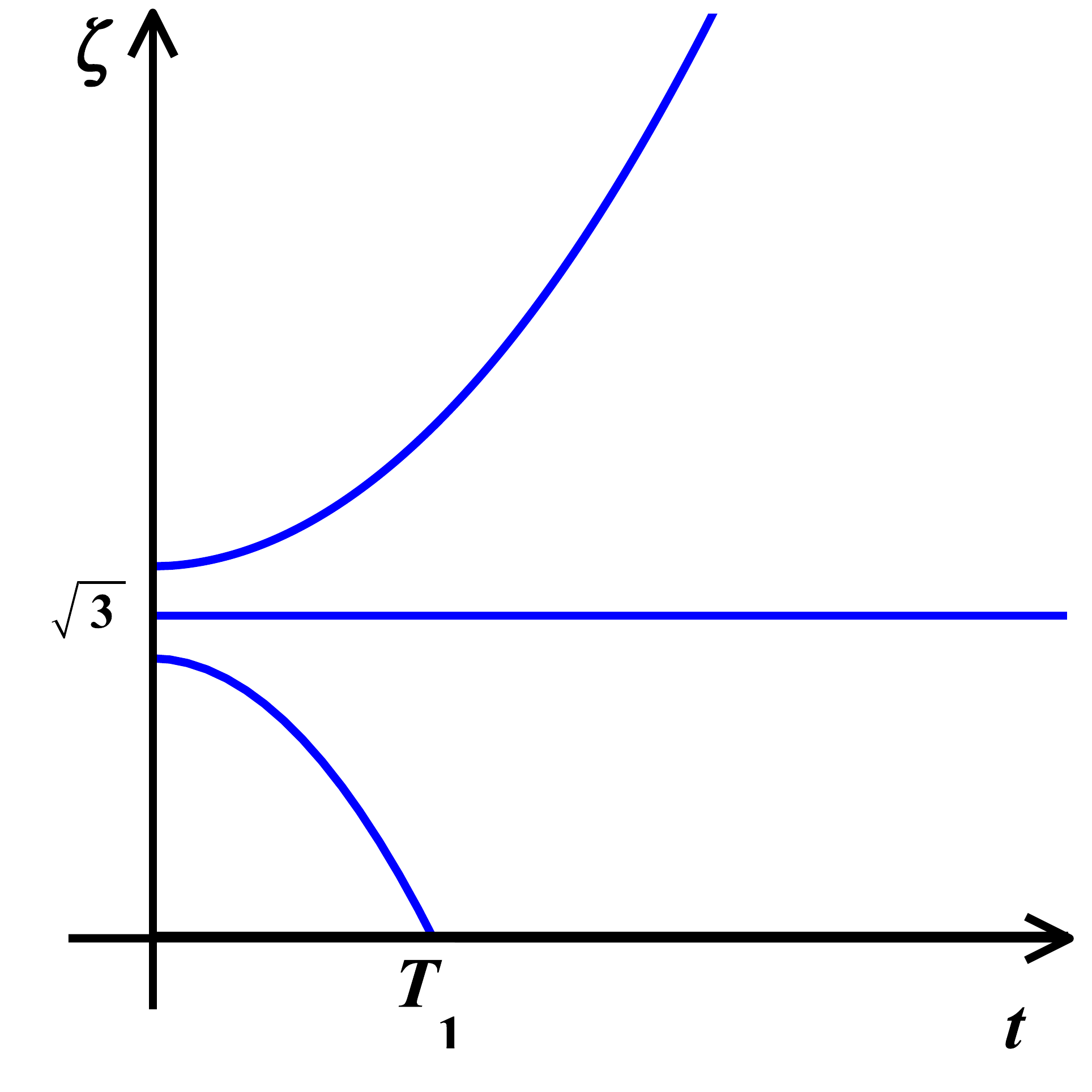}
    \caption{Solution $\zeta (t)$ for different values of $\zeta (0)$}
    \label{fig:1D_2abra}
\end{center}
\end{figure}

Substituting it into $\left( \ref{11}\right) $ we have%
\begin{equation*}
\frac{d}{dt}\underset{\mathbf{0}}{\overset{\zeta(t)}{\int
}}\frac{1}{8}x^{4}dx
=\underset{\mathbf{0}}{\overset{\zeta (t)}{\int }}x^{2}dx-\underset{\mathbf{0}}{%
\overset{\zeta(t)}{\int }}dx
\end{equation*}
i.e., it is equivalent to
\begin{equation*}
 \frac{1}{8}\zeta^{3}(t)\zeta^{\prime }(t)
=\frac{1}{3}\zeta^{2}(t)-1.
\end{equation*}
The solution to this differential equation is given implicitly by
\[
t=\frac{9}{16}\left[ \frac{1}{3}\zeta
^{2}(t)-\frac{1}{3}\zeta^{2}(0)+\ln \left\vert
\frac{\frac{1}{3}\zeta ^{2}(t)-1}{\frac{1}{3}\zeta
^{2}(0)-1}\right\vert \right]
\]%
(see Fig.\ref{fig:1D_2abra}) and $\zeta (t)$ vanishes at
\begin{equation*}
T_1=\frac{9}{16}\left[ -\ln \left\vert \frac{1}{3}\zeta ^{2}(0)-1\right\vert -\frac{1%
}{3}\zeta ^{2}(0)\right]
\end{equation*}
if $\zeta (0)< \sqrt{3}$. Note, that $T_1 \to \infty$ as $\zeta (0)
\to \sqrt{3}$.

If the initial wavelength $\lambda =2\zeta (0)$ is larger than
$2\sqrt {3}$, then the amplitude and the period growth without
limit. In this case, the period behaves like $\sqrt {t}$ (typical
behavior) and  the amplitude behave like $t$ as $t$ tends to $\infty
$.

The case $\lambda(0)= 2\sqrt{3}$  leads to a stationary periodic
solution, i.e., $\lambda(t)=2\sqrt{3},$ for all $t.$

It turns out that property (\ref{11}), which played a crucial role
in the time behavior of solutions, still valid even if the CKPZ term
is not present. At  first sight, we may deduce that the CKPZ term,
which acts at small scales, has no effect on the time-behavior of
the typical length.

\subsection{Two-dimensional problem}

Next, we investigate the nonlinear deterministic equation in two
dimensions. Let us take the conserved Kuramoto-Sivashinsky equation
as growth equation in the form
\begin{equation}
u_{t}+\Delta \left( u+\Delta u+r \left\vert \nabla u\right\vert
^{2}\right) =0  \tag{CKS-2}.  \label{CKS-2}
\end{equation}%

We suppose that $\mathbf{x}\in \mathbf{R}^{2},$ and $u$ is a $C^{2}$
smooth function. Multiplying (\ref{CKS-2})  with $u$ and integrating
twice one gets
\begin{equation}
\begin{split}
\frac{d}{dt}\frac{1}{2}\underset{\Omega }{\int }u^{2}dx~dy=\underset{\Omega }%
{\int }\left\vert \nabla u\right\vert ^{2}dx~dy \\ -\underset{\Omega }{\int }%
\left( \Delta u\right) ^{2}dx~dy-r \underset{\Omega } \int
\left\vert \nabla u\right\vert ^{2}\Delta u dx~dy  \label{2D}
\end{split}
\end{equation}%
for any rapidly decreasing solution $u$.

If $u$ represents the mound like growth of the form  of $u=C \left(
x^{2}+y^{2}\right) $ with some parameter $C<0,$ then one obtains
from the partial differential equation (\ref{CKS-1}) that
\[
C=-\frac{1}{4r },
\]%
i.e.,
\[
u=-\frac{1}{4r }\left( x^{2}+y^{2}\right)
\]%
and%
\[
u(x,y,t)=\left\{
\begin{array}{c}
-\frac{1}{4 r}\left( x^{2}+y^{2}\right) \text{ if }x^{2}+y^{2}<\chi ^{2}(t) \\
0\text{ \ \ \ if \ \ \ \ }x^{2}+y^{2}\geq \chi ^{2}(t)%
\end{array}%
\right.
\]%
with some function $\chi $. Then $\max \vert u\vert =\chi ^{2}/{4r}$. Taking the integrals for $\Omega =B\left( 0,\chi \right) $\ in (\ref%
{2D}) we obtain the differential equation for $\chi $
\begin{equation}
\frac{1}{8}\chi ^{3}(t)\chi ^{\prime }(t) =\frac{1}{2}\chi
^{2}(t)-2.
\end{equation}

By integration one gets the solution as
\begin{equation}
t =\frac{1}{2}\left[ \frac{1}{4}\chi ^{2}-\frac{1}{4}\chi
^{2}(0)+\ln \left\vert \frac{\frac{1}{4}\chi ^{2}-1}{\frac{1}{4}\chi
^{2}(0)-1}\right\vert \right]
\end{equation}
\begin{figure}
\begin{center}
    \includegraphics[keepaspectratio, width=4.5cm]{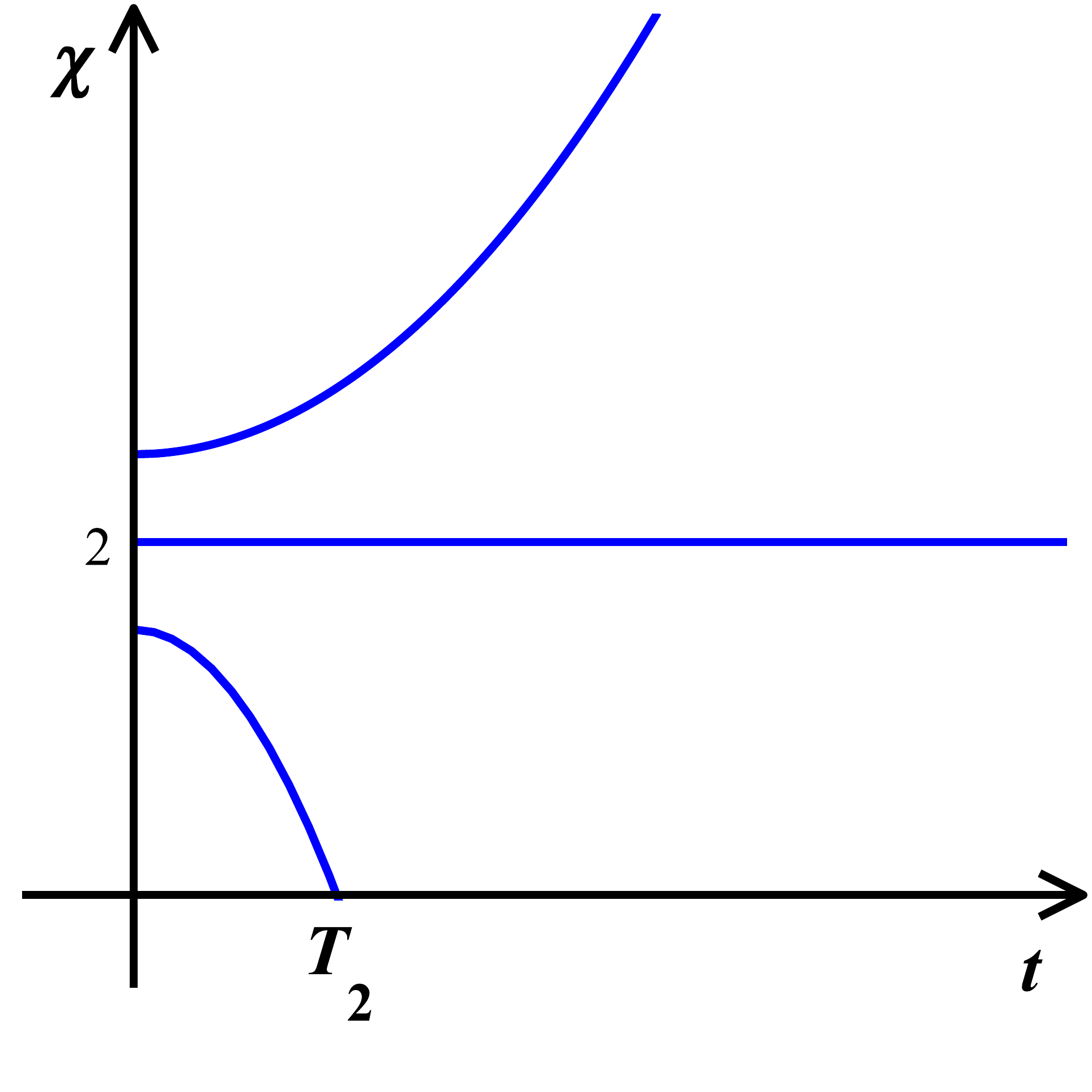}
    \caption{Function $\chi (t)$ for different initial values of $\chi (0)$}
    \label{fig:1D_3abra}
\end{center}
\end{figure}
If $\chi (0)<2$, then $\chi (t)$ collapses at finite $T_2,$ see Fig.\ref{fig:1D_3abra}, where
\begin{equation}
T_2 =\frac{1}{2}\left[ - \ln \left\vert \frac{1}{4}\chi ^{2}(0)-1
\right\vert -\frac{1}{4}\chi ^{2}(0)\right].
\end{equation}
For $\chi (0) > 2$, the period and amplitude for large initial data
behave like in the one-dimensional case, while if $\chi (0) =2$ we
obtain a stionnary periodic solution.

\subsection{The effect of the KPZ term}

Next, we study analytically the effect of the nonlinear term in
(\ref{CKS-2}).
As above we consider the two and one-dimensional cases separately;
in two-dimension
\begin{equation}
u_{t}=-\Delta \left( \left\vert \nabla u\right\vert ^{2}\right)
\label{2D-1} \end{equation} and in one-dimension
\begin{equation}
u_{t}=-\left( u_{x}\right)^2 _{xx}. \label{1D-1}
\end{equation}

With substitution $v=u_{x}$ to equation (\ref{1D-1}) one gets
\[
v_{t}=-(v^{2})_{xxx}.
\]
Self-similar solution for $v$ can be searched in the form
\[
v(x,t)=(T_3-t)^{\alpha }g\left( x\left( T_3-t\right) ^{-\beta
}\right),  \ \
\]%
for some $T_3$. Then
the differential equation (\ref{1D-1}) takes the form of
\[
\left( g^{2}\right) ^{\prime \prime \prime }-\alpha g+\beta \eta
g^{\prime }=0
\]%
 when
\begin{equation}
 -\alpha +3\beta =1  \label{cond}
\end{equation}
for $g=g(\eta)$ with $\eta=x\left( T_3-t\right)^{-\beta }$.
Physically, as we have conserved equation, we must have $\alpha
+\beta =0$, and therefore with (\ref{cond}) one gets $-\alpha =\beta
=1/4.$ Then
\begin{equation}
\frac {1}{4} \eta g+\left( g^{2}\right) ^{\prime \prime }=K_1,
\label{3}
\end{equation}
where $K_1$ is a constant.

(i.) If $K_1=0$ and $g(\eta )=C \eta ^3$ then for (\ref{3}) we get
that $g=-\eta ^3/120$ and
\begin{equation*}
\begin{split}
v=-\frac {1}{120}\left( T_3-t\right) ^{-\frac{1}{4}}\left( x\left( T_3-t\right) ^{-\frac{1}{4}%
}\right) ^{3} \\ =  -\frac {1}{120}\frac{x^{3}}{T_3-t}.
\end{split}
\end{equation*}

Therefore solution $u$ reads as
\begin{equation}
u=-\frac {1}{480}\frac{x^{4}}{T_3-t}+K_2  \label{13} \end{equation}
with some constant $K_2$.

It is easy to see from (\ref{13}) that the wavelength $\lambda =2
(480 K_2(T_3-t))^{1/4}$ collapses at finite time.

\begin{figure}
\begin{center}
    \includegraphics[keepaspectratio, width=7cm]{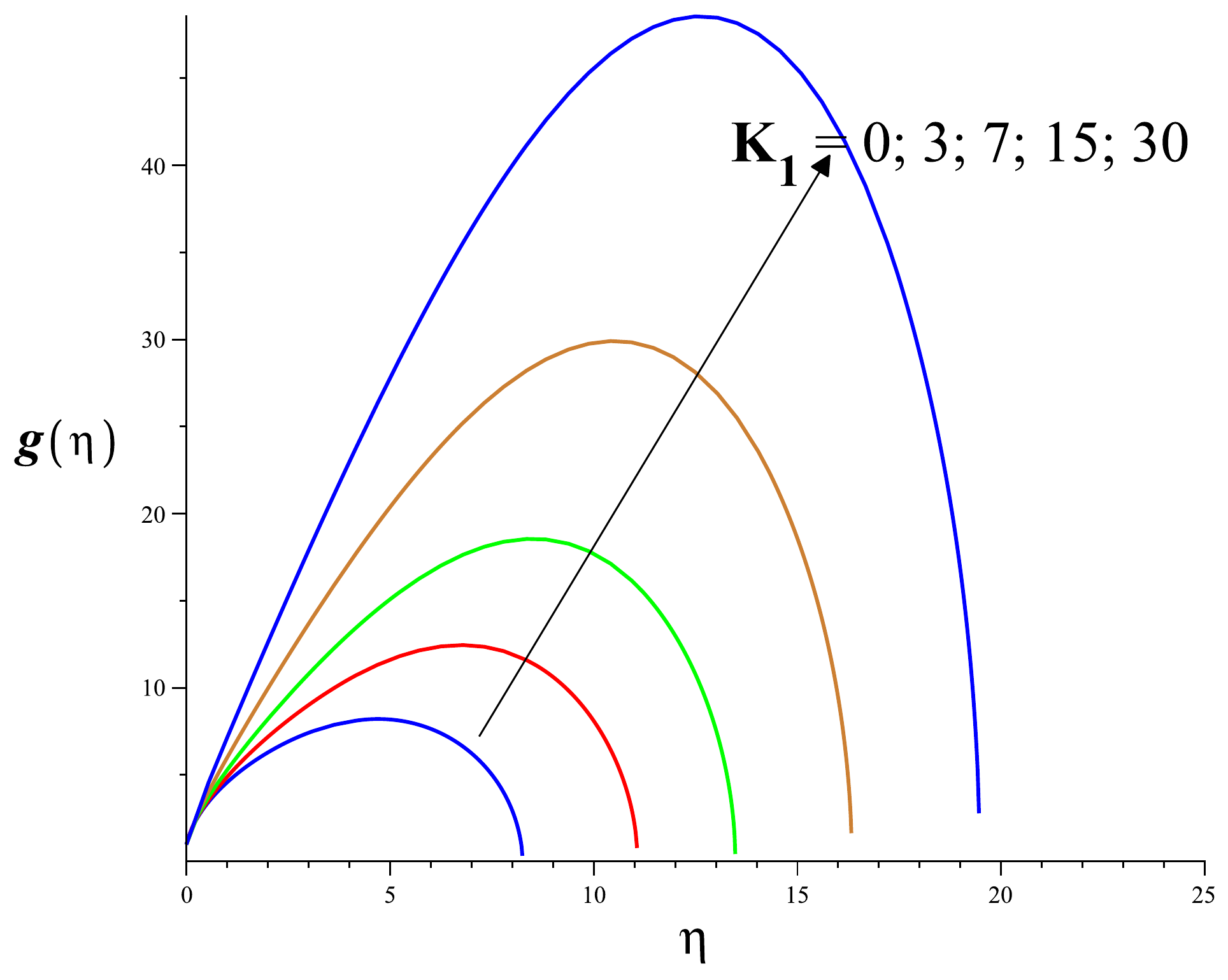}
    \caption{Solutions to (\ref{3}) for different values of $K_1$}
    \label{fig:1D_5abra}
\end{center}
\end{figure}

(ii.) The numerical solutions to equation (\ref{3}) with initial
conditions  $ f(0)=1,\text{ \ }f^{\prime }(0)=10 $  can be obtained
(see Fig.\ref{fig:1D_5abra}).

(iii.) At $\eta =\eta_k$ the differential equation ($\ref{3}$)
behaves like
\begin{equation}
gg^{\prime \prime }+g^{\prime 2}=0 \label{3-0}.
\end{equation}
For this equation the solution
\[
g\sim \sqrt{\eta _{k}-\eta }
\]%
has vertical asymptote at $\eta _{k}.$ Hence,
\begin{equation}
v\sim \left( T_3-t\right) ^{-\frac{1}{4}}\sqrt{\eta _{k}-x \left(
T_3-t\right) ^{-\frac{1}{4}} }
\end{equation}
and
\begin{equation}
u\sim \left[K_3 - \frac{2}{3} \left(\eta_k- x \left( T_3-t\right)
^{-\frac{1}{4}}\right)^\frac{3}{2} \right ]_+
\end{equation}
with a constant $K_3$. Then, $u$ collapses at finite time.

In case of two-dimensional problem (\ref{2D-1}), the solution $u$ is
searched in the form
\begin{equation}
u(x,y,t)=\left[ a(T_4-t)-b(x^2+y^2)\right]_+
\end{equation}
for some constants $a$, $b$ and $T_4$. Substituting $u$ into
(\ref{2D-1}), we obtain $a=16b^2$ and
\begin{equation}
u(x,y,t)=b(T_4-t)\left[ 16b-\frac{x^2+y^2}{T_4-t}\right]_+,
\end{equation}
this is precisely a similarity solution, which shows that both the
amplitude $A=b(T_4-t)$ and the wavelength $\lambda
=4\sqrt{b}\sqrt{T_4-t}$ decrease with time and vanish at finite time
$T_4$. This may give some light to the qualitative effect of the
CKPZ term on physical properties of the CKPZ equation.  The aim of
the next section is to investigate the effect of the KPZ term by
using numerical solutions.

\section{Numerical results}

Both the one-dimensional and two-dimensional generalized
Kuramoto-Sivashinsky equation are solved with periodic boundary
conditions using Fourier spectral collocation in space and the
fourth order Runge-Kutta exponential time differencing scheme for
time discretization.

\subsection{One-dimensional case}

First, we investigate the one-dimensional generalized
Kuramoto-Sivashinsky equation
\begin{equation}
\begin{split}
   u_t = -u_{xx} - u_{xxxx} + (u_x)^2- r\left((u_x)^2\right)_{xx}, \\ x\in(x_1,x_2),\ t>0
   \label{eq:ourIBVP-1}
\end{split}
\end{equation}
with periodic boundary condition
\begin{equation}
   u(x_1) = u(x_2)
   \label{eq:ourIBVP_BC}
\end{equation}
for some points $x_1$ and $x_2$ and with initial condition
\begin{equation}
   u(x,0) = u_0(x).
   \label{eq:ourIBVP_IC}
\end{equation}
Let us discretize function $u$ of \eqref{eq:ourIBVP-1} in $x$ for
$N$ equidistant points and then take its discrete Fourier transform.
Taking the time derivative component-wise and using formula for the
transform of the derivatives, we get from \eqref{eq:ourIBVP-1} a
system of ordinary differential equations.
Function $({u}_x)^2$ is evaluated pseudospectrally in the Fourier
space.

For temporal discretization of the system we apply the exponential
time differencing method ETDRK4 scheme, a fourth order Runge-Kutta
time stepping introduced by Cox and Matthews in \cite{Cox2002}. The
program for solving the IBVP
\eqref{eq:ourIBVP-1}--\eqref{eq:ourIBVP_IC} is created both in
MATLAB (version R2011a) and in C++ using the ArrayFire library
(version 3.0.1, build 17db1c9).
In MATLAB, the default double precision type is used to be able to
exploit the precision of the spectral method. In ArrayFire, the
variables are declared as real and complex double precision types
(\texttt{f64} and \texttt{c64} types, respectively).

\begin{figure}
\begin{center}
    \includegraphics[keepaspectratio, width=7cm]{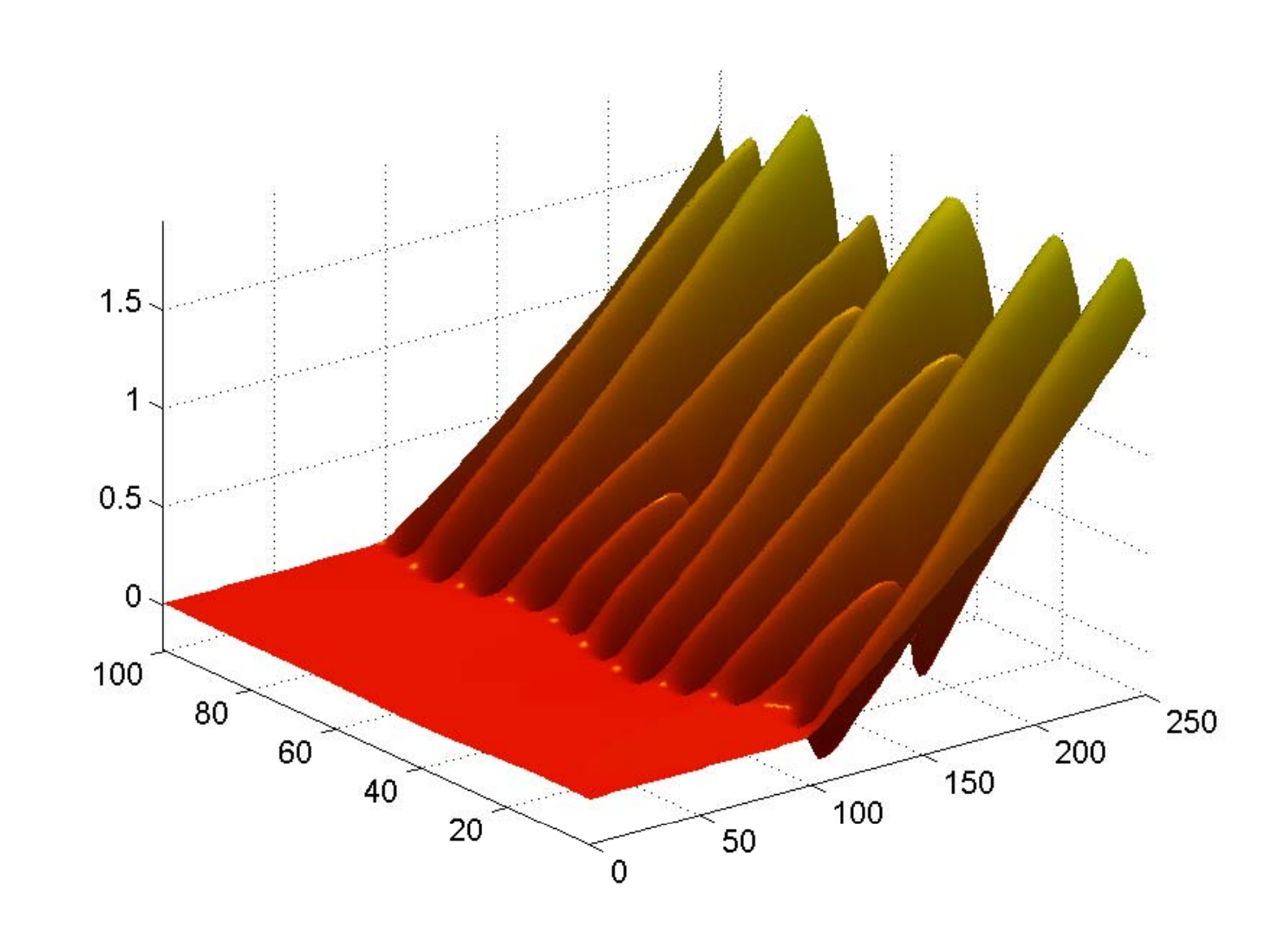}
    \caption{Solution to \eqref{eq:ourIBVP-1}, \eqref{eq:IC_1D} for $r=10$ and $A(0)=0.01$}
    \label{fig:1D_r=10}
\end{center}
\end{figure}

In the one-dimensional case the initial condition  used is of the
form
\begin{equation}
   u(x,0) =A(0)  \cos\frac{x}{16}  \left( 1+\sin\frac{x}{16}
   \right),
   \label{eq:IC_1D}
\end{equation}
where $A$ is a positive constant. Figures
\ref{fig:1D_r=0,01}-\ref{fig:1D_r=10} show the numerical solution
for differently chosen $r$ with parameters $N=256$, $\Delta t=1/100$
on $x\in[0,32\pi]$, $t\in[0,250]$.
 For contour
integration with trapezoidal rule, $M=32$ is applied.
\begin{figure}
\begin{center}
    \includegraphics[keepaspectratio, width=7cm]{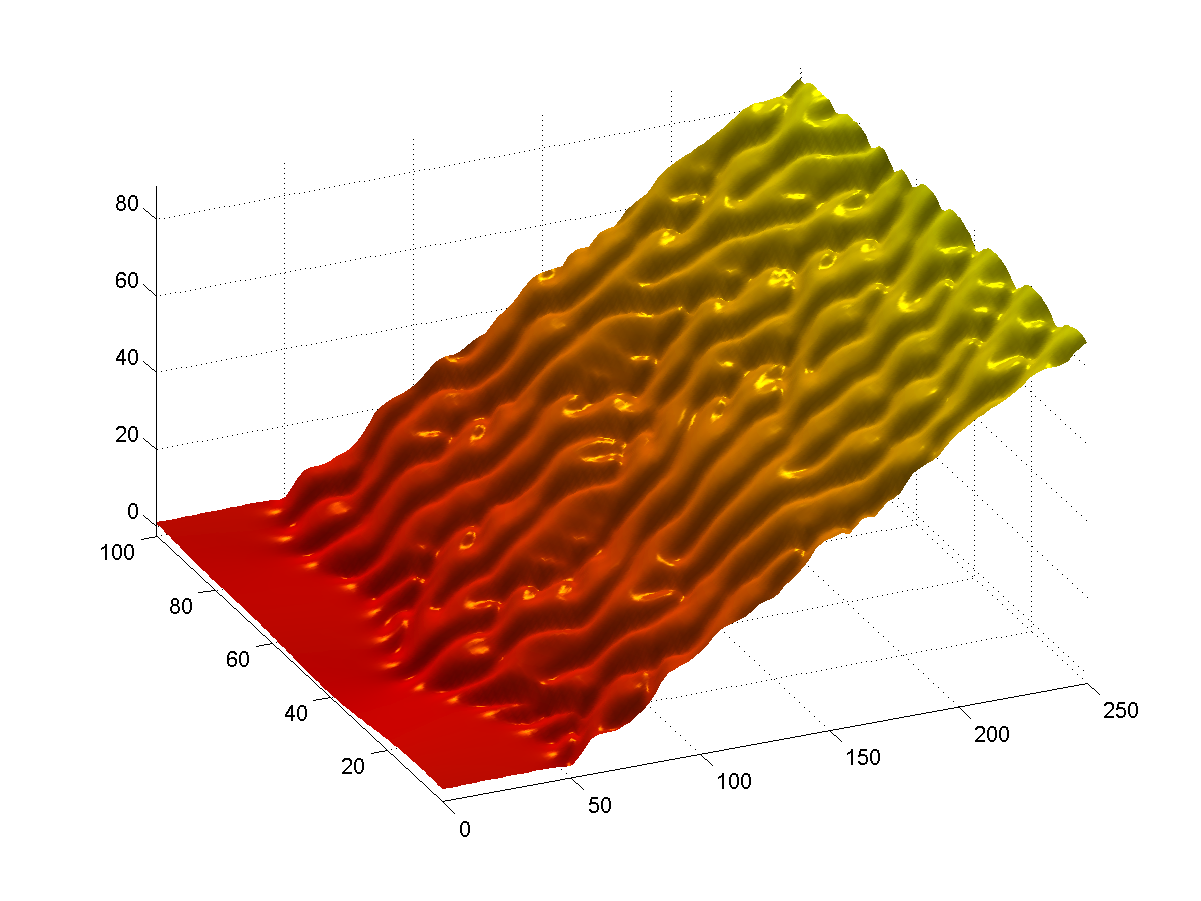}
    \caption{Solution to \eqref{eq:ourIBVP-1}, \eqref{eq:IC_1D} for $r=0.01$ and $A(0)=1$}
    \label{fig:1D_r=0,01}
\end{center}
\end{figure}
\begin{figure}
\begin{center}
    \includegraphics[keepaspectratio, width=7cm]{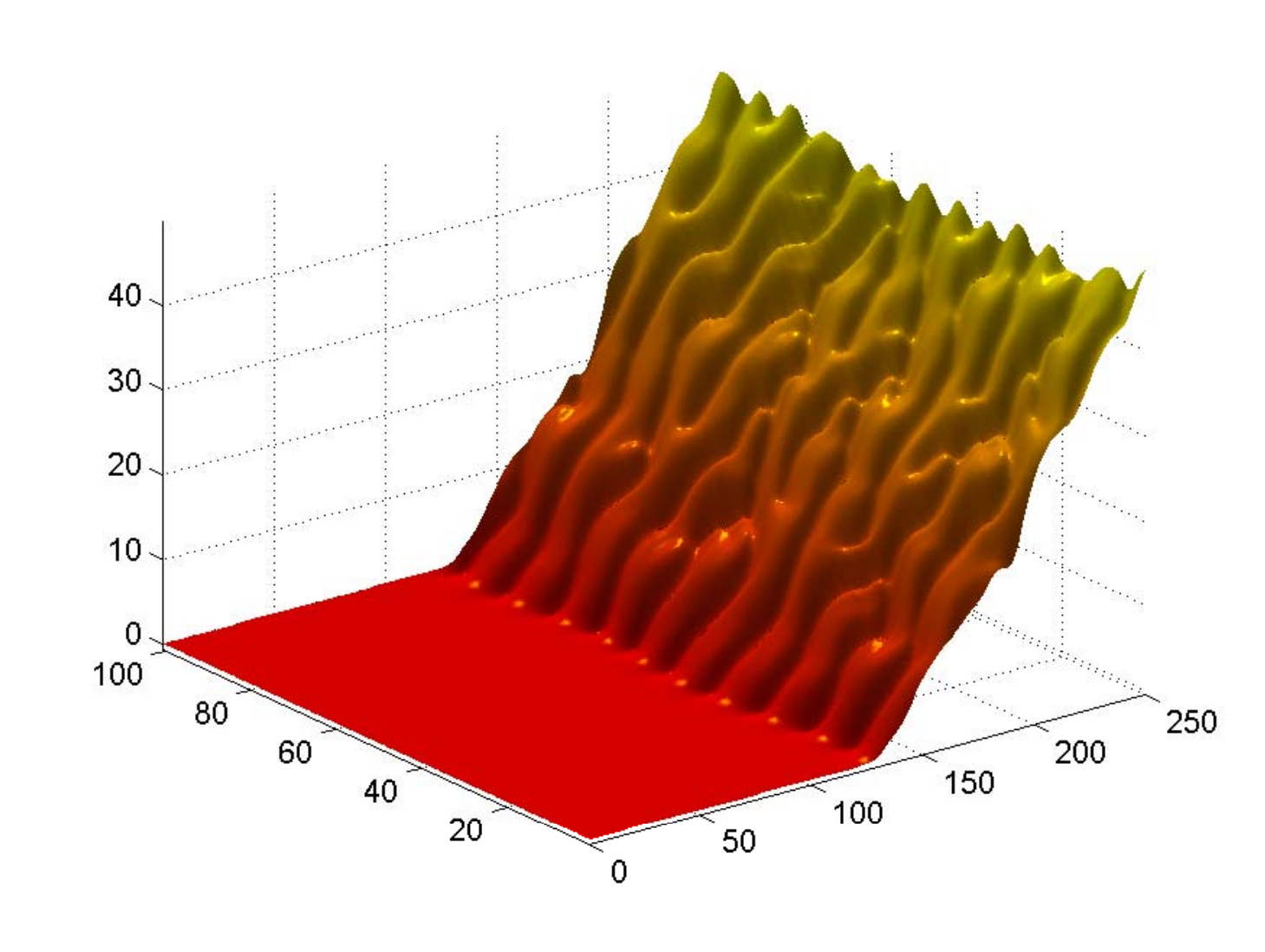}
    \caption{Solution to \eqref{eq:ourIBVP-1}, \eqref{eq:IC_1D} for $r=0.01$ and $A(0)=0.01$}
    \label{fig:1D_r=0,01-0,01}
\end{center}
\end{figure}
\begin{figure}
\begin{center}
    \includegraphics[keepaspectratio, width=7cm]{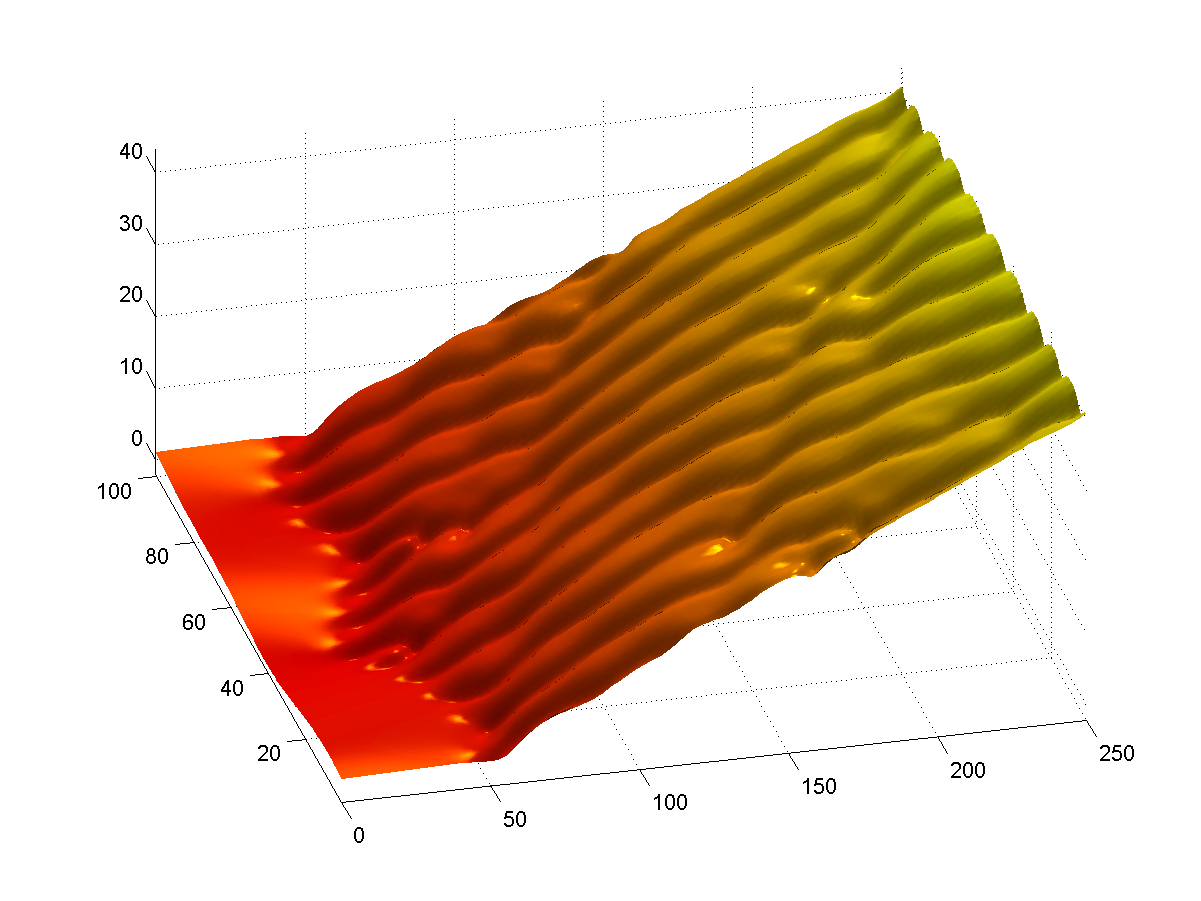}
    \caption{Solution to \eqref{eq:ourIBVP-1}, \eqref{eq:IC_1D} for $r=0.5$ and $A(0)=1$}
    \label{fig:1D_r=0,5}
\end{center}
\end{figure}

The following consequences can be drawn from the figures. First, the
increase of $r$ causes the solution function $u$ grow less, meaning
that the accumulation of atoms on the surface is far less
significant that for smaller $r$. Second, the chaotic nature of the
solution emerges later in time when $r$ is large. Third, the
accumulation of the atoms to the surface starts later with smaller
value of $A$ (see Fig. \ref{fig:1D_r=0,01} and
\ref{fig:1D_r=0,01-0,01}). It is not seen in the figures, but the
solution became bounded for a longer time interval when $r$ was
small. It can either be a numerical issue or the exact solution
itself blows up at finite time; it needs further investigations.

Efficient MATLAB and ArrayFire codes were written to tackle the
initial value problem numerically. The simulated results show
physically meaningful characteristics and are similar to the results
obtained with microscopic measurements.

\subsection{Two-dimensional case}

In case of the 2D problem, equation
\begin{equation}
   u_t + \Delta\left( u+\Delta u + r|\nabla u|^2 \right) = 0, \quad (x,y)\in(a,b)^2
   \label{eq:ourIBVP_2D}
\end{equation}
is solved with different initial conditions on $(x,y)\in[0,32\pi]^2$
using parameters $r=10$, $N = 256$, $\Delta t = 1/100$, $M=32$. The
following initial conditions are considered
\begin{gather}
   u(x,y,0) = 0.01\left| \sin\frac{x^2+y^2}{16} \right|,
   \label{eq:IC1_2D} \\
   u(x,y,0) = 0.01\left( \sin\frac{x^2+y^2}{16} + \left| \sin\frac{x^2+y^2}{16} \right|
   \right),
   \label{eq:IC2_2D} \\
   u(x,y,0) = 0.1 \sin\frac{x^2}{16}\cos\frac{y^2}{16}.
   \label{eq:IC3_2D}
\end{gather}
The two-dimensional problem \eqref{eq:ourIBVP_2D} is solved
numerically with the initial condition
\begin{equation}
    u(x,y,0) = u_0(x,y).
    \label{eq:ourIBVP2D_IC}
\end{equation}

The implementation of the two-dimensional case is similar to the
one-dimensional one, because the two-dimensional discrete Fourier
transform also results in decoupled ordinary differential equations.
The only difference is that the unknowns now constitute a matrix
instead of a vector as in the one-dimensional case. The temporal
discretization schemes apply to scalar equations, therefore we can
take the formulas component-wise as we did before.

\begin{figure}
\begin{center}
    \includegraphics[keepaspectratio, width=8cm]{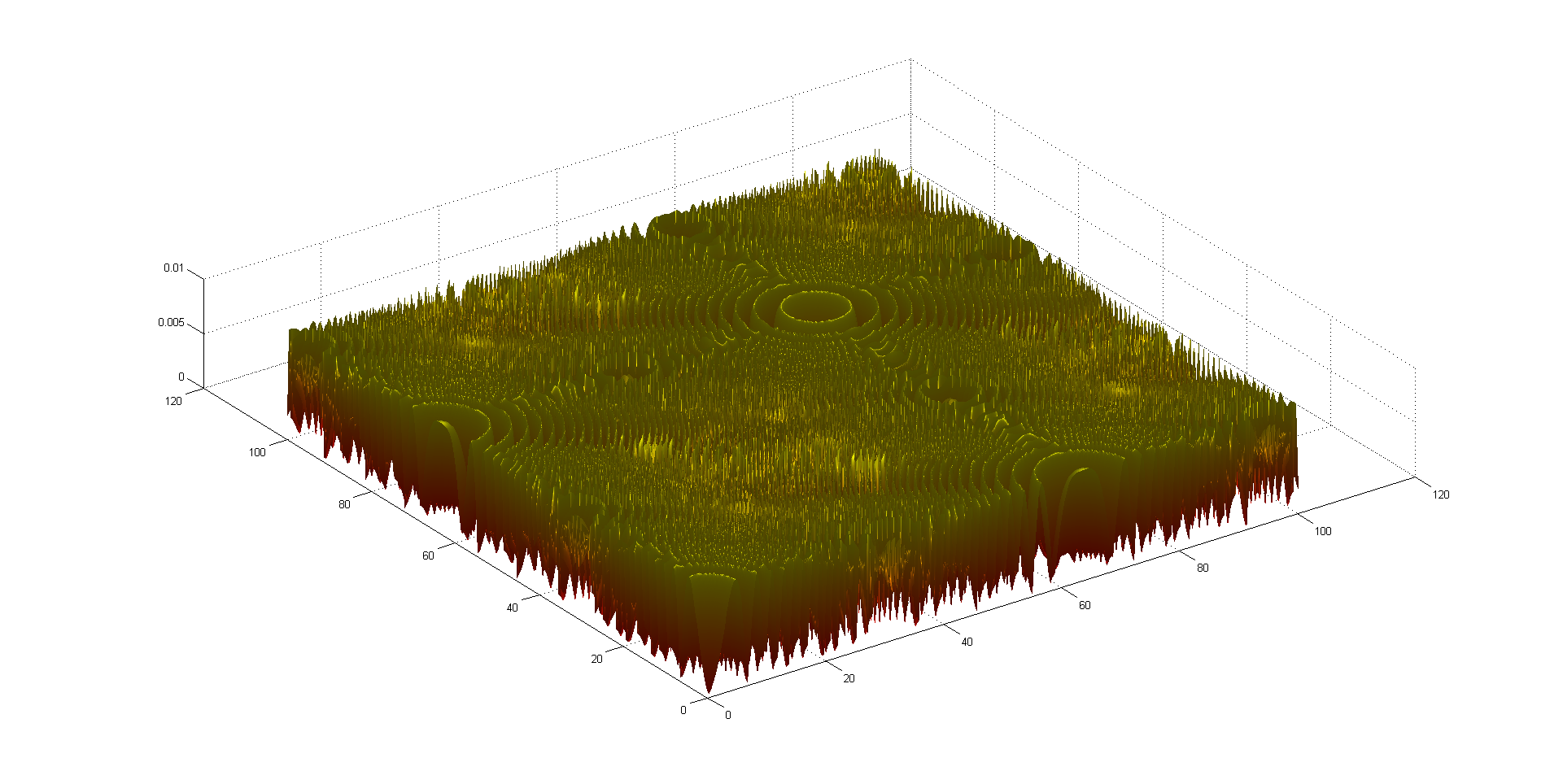}
    \caption{Solution to \eqref{eq:ourIBVP_2D}, \eqref{eq:IC1_2D} for $r=10$ at $t=0$}
    \label{fig:IC1_t=0}
\end{center}
\end{figure}

\begin{figure}
\begin{center}
    \includegraphics[keepaspectratio, width=8cm]{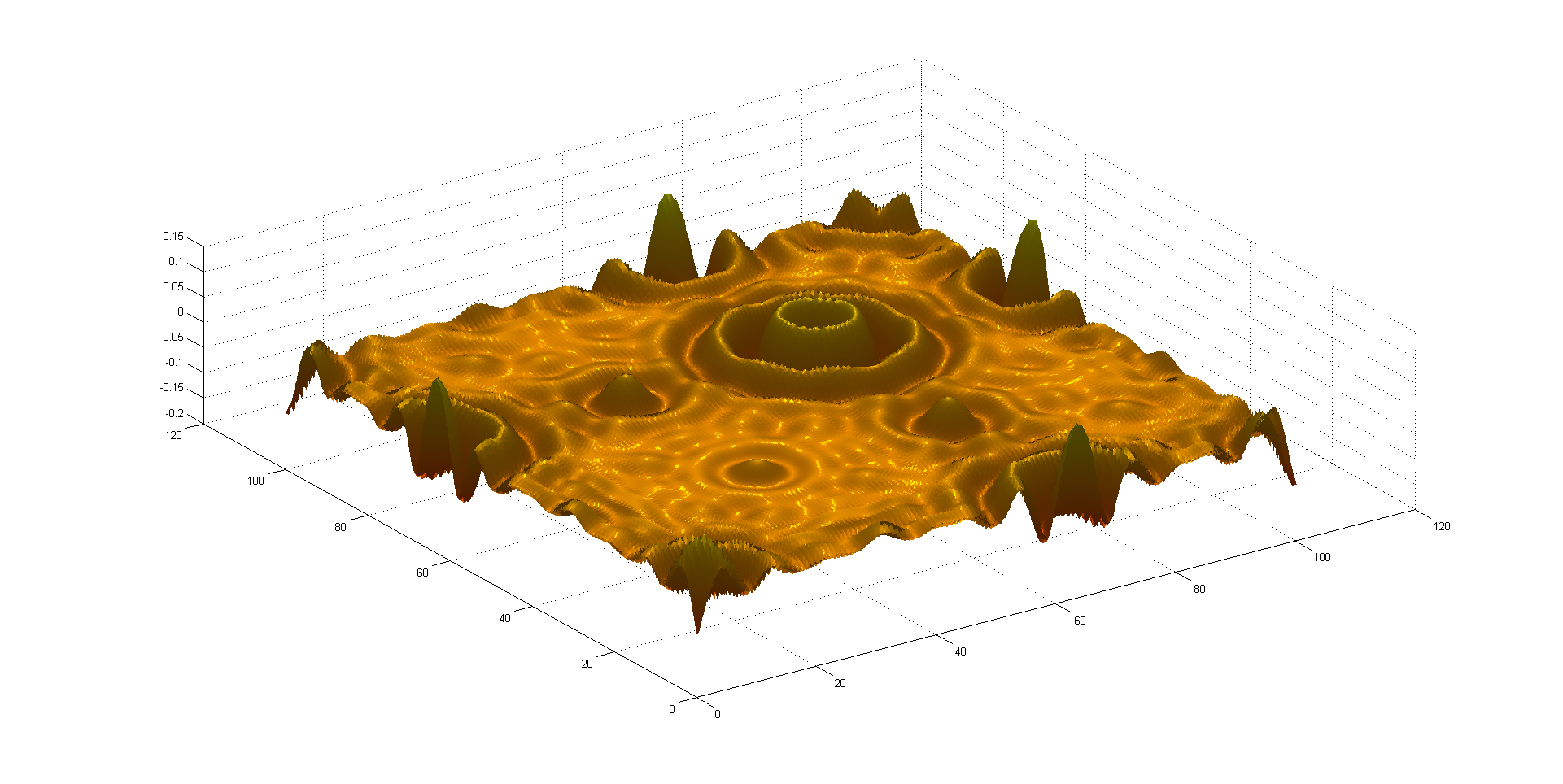}
    \caption{Solution to \eqref{eq:ourIBVP_2D}, \eqref{eq:IC1_2D} for $r=10$ at $t=20$}
    \label{fig:IC1_t=20}
\end{center}
\end{figure}

\begin{figure}
\begin{center}
    \includegraphics[keepaspectratio, width=8cm]{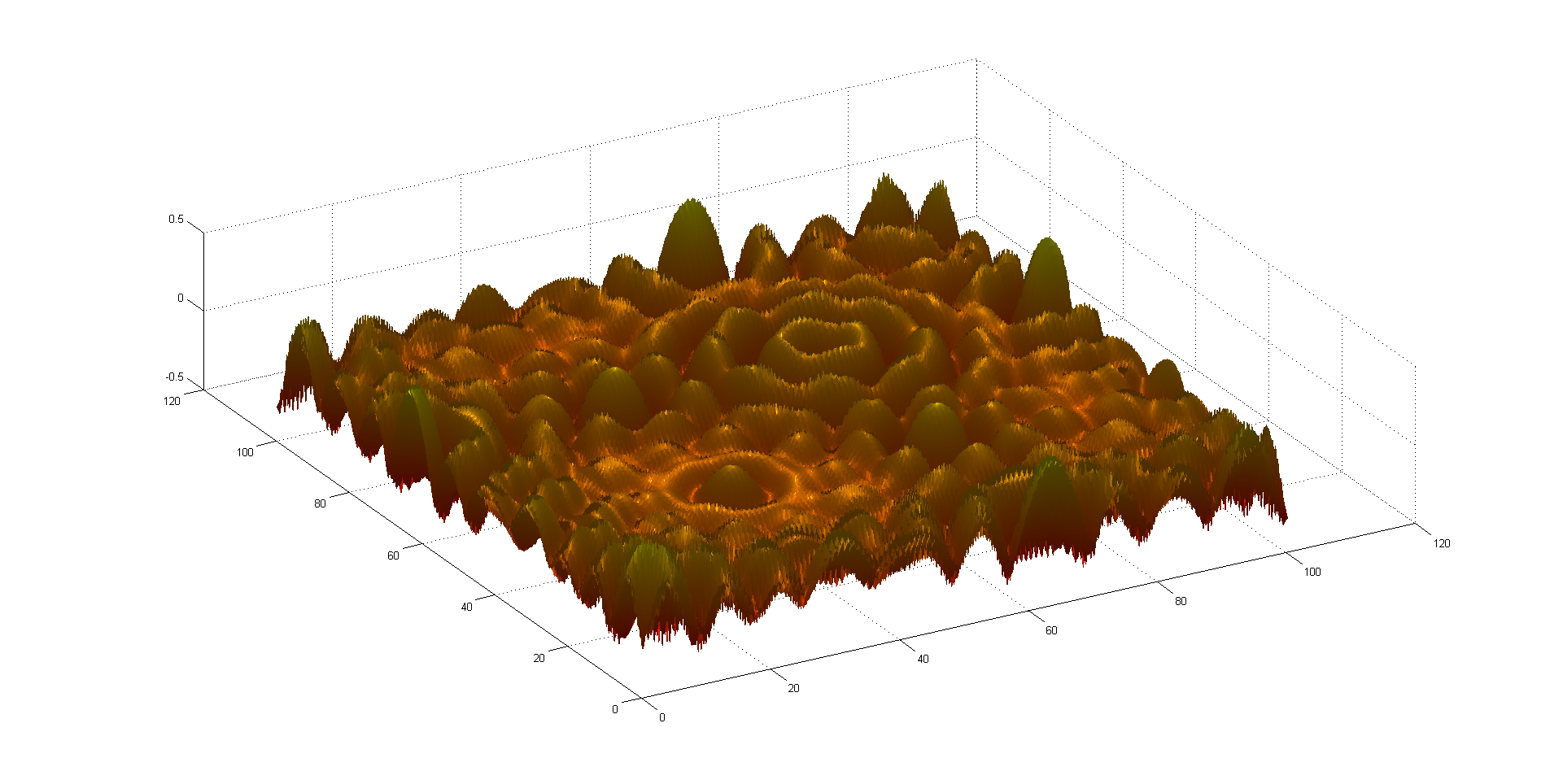}
    \caption{Solution to \eqref{eq:ourIBVP_2D}, \eqref{eq:IC1_2D} for $r=10$ at $t=30$}
    \label{fig:IC1_t=30}
\end{center}
\end{figure}

\begin{figure}
\begin{center}
    \includegraphics[keepaspectratio, width=8cm]{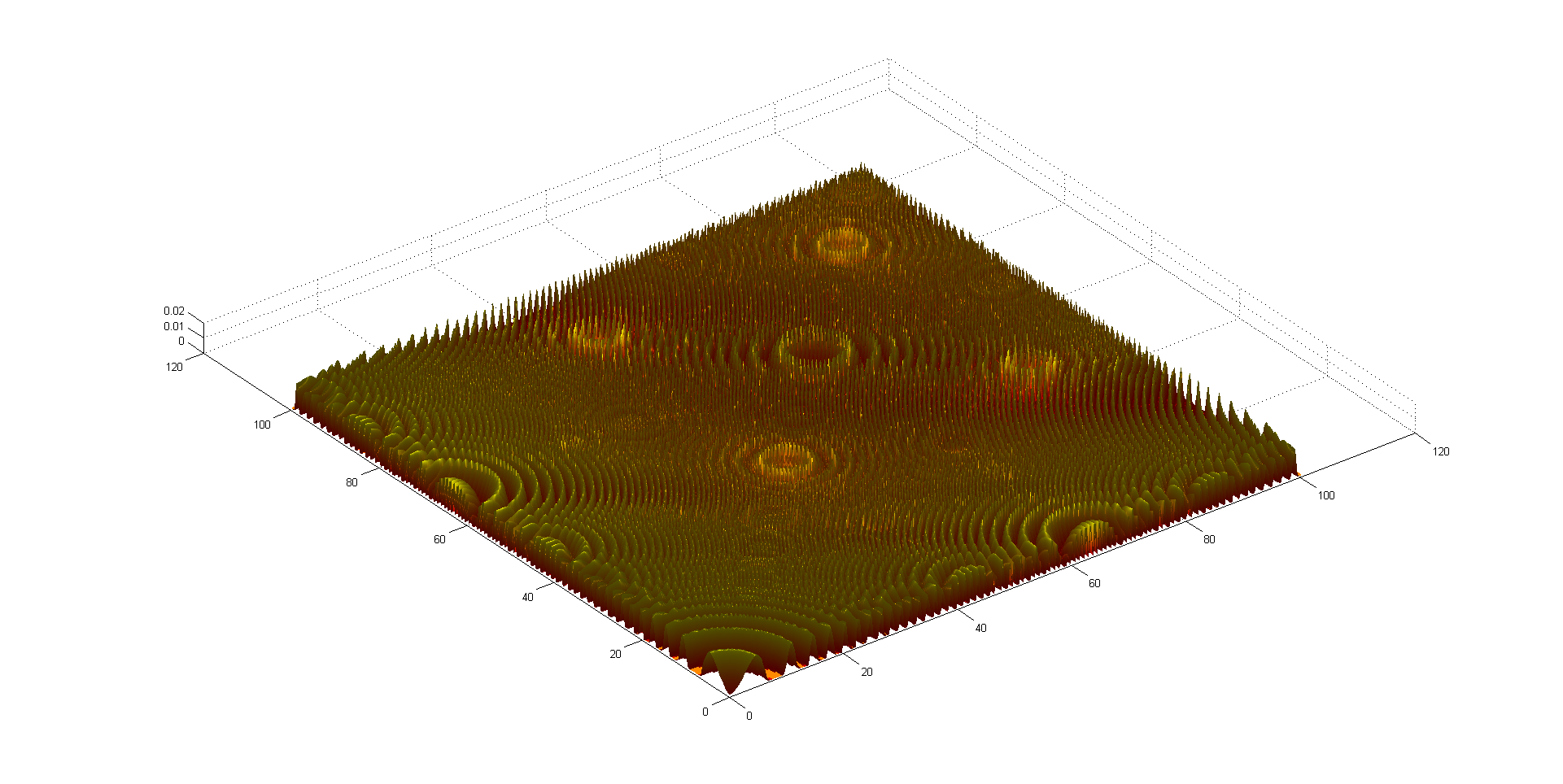}
    \caption{Solution to \eqref{eq:ourIBVP_2D}, \eqref{eq:IC2_2D} for $r=10$ at $t=0$}
    \label{fig:IC2_t=0}
\end{center}
\end{figure}

\begin{figure}
\begin{center}
    \includegraphics[keepaspectratio, width=8cm]{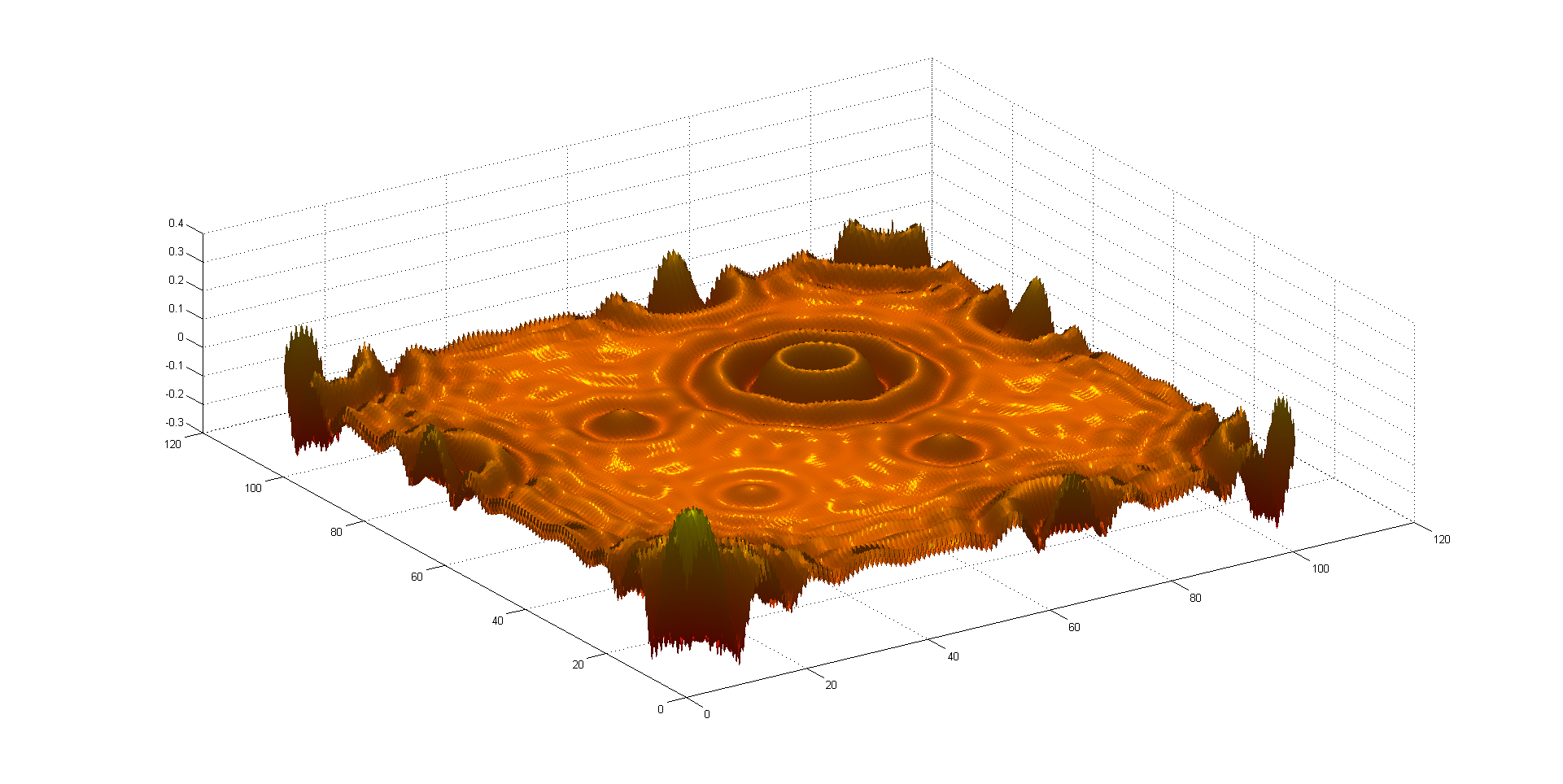}
    \caption{Solution to \eqref{eq:ourIBVP_2D}, \eqref{eq:IC2_2D} for $r=10$ at $t=20$}
    \label{fig:IC2_t=20}
\end{center}
\end{figure}

\begin{figure}
\begin{center}
    \includegraphics[keepaspectratio, width=8cm]{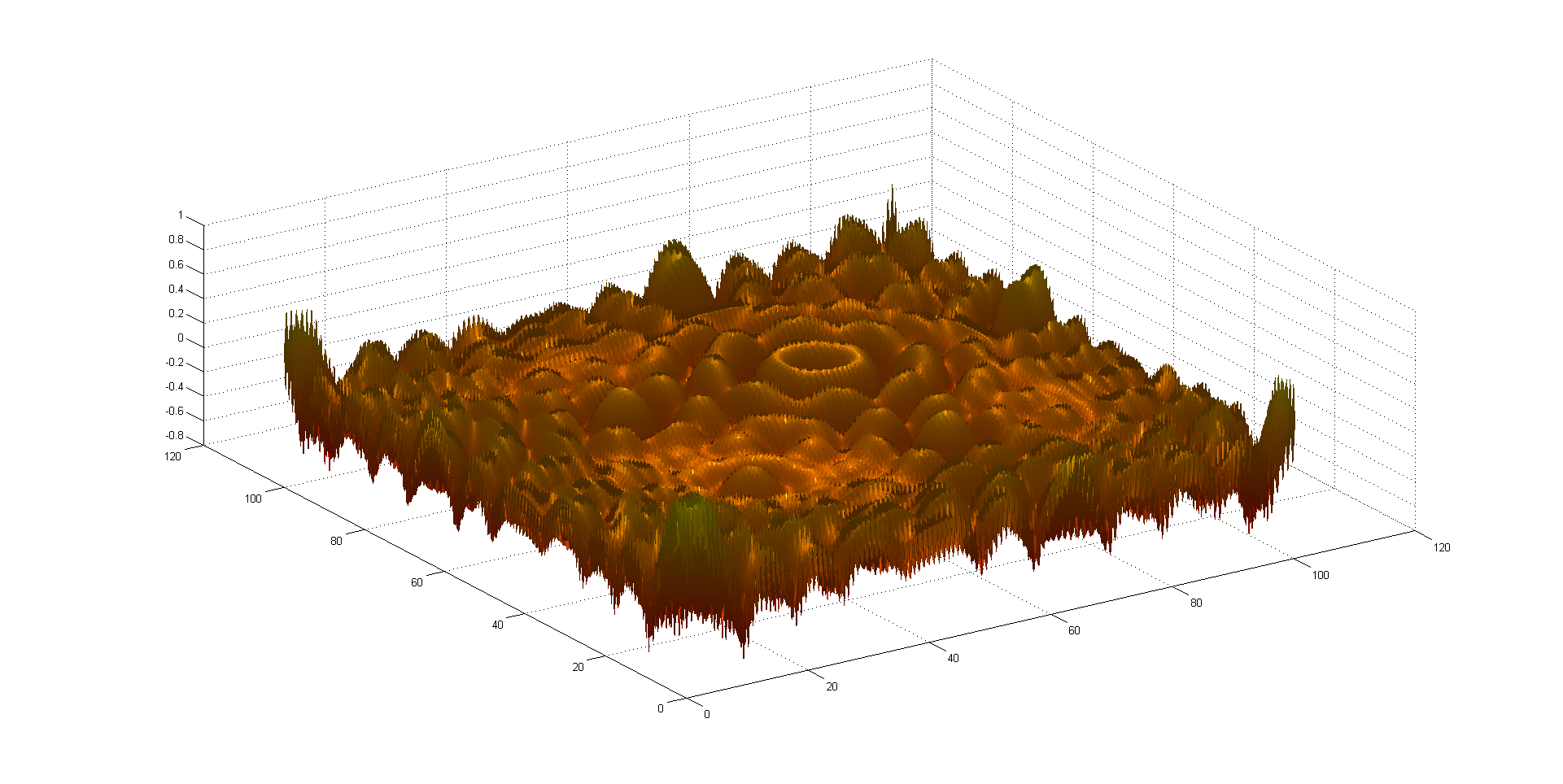}
    \caption{Solution to \eqref{eq:ourIBVP_2D}, \eqref{eq:IC2_2D} for $r=10$ at $t=30$}
    \label{fig:IC2_t=30}
\end{center}
\end{figure}

\begin{figure}
\begin{center}
    \includegraphics[keepaspectratio, width=8cm]{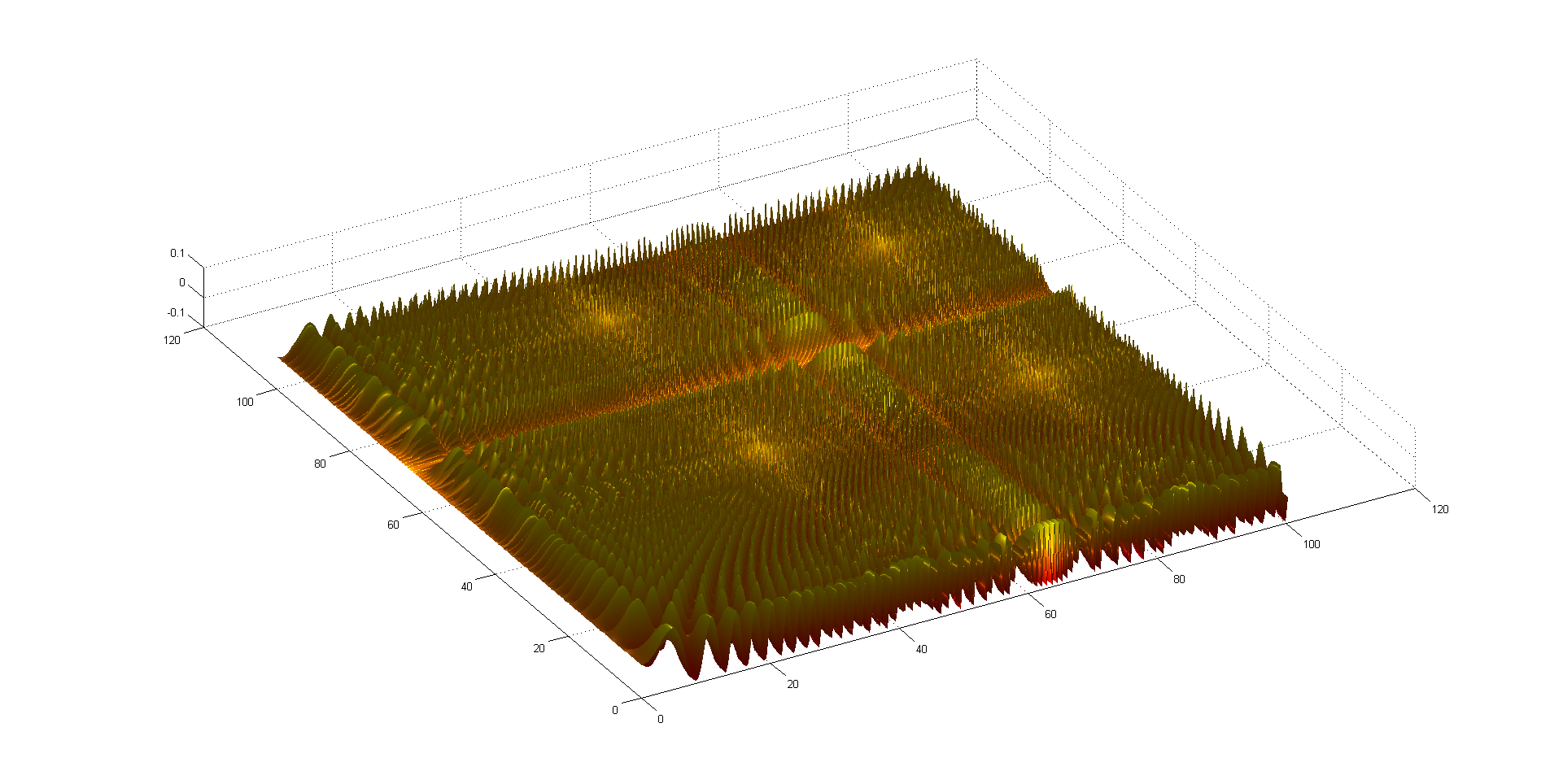}
    \caption{Solution to \eqref{eq:ourIBVP_2D}, \eqref{eq:IC3_2D} for $r=10$ at $t=0$}
    \label{fig:IC3_t=0}
\end{center}
\end{figure}

\begin{figure}
\begin{center}
    \includegraphics[keepaspectratio, width=8cm]{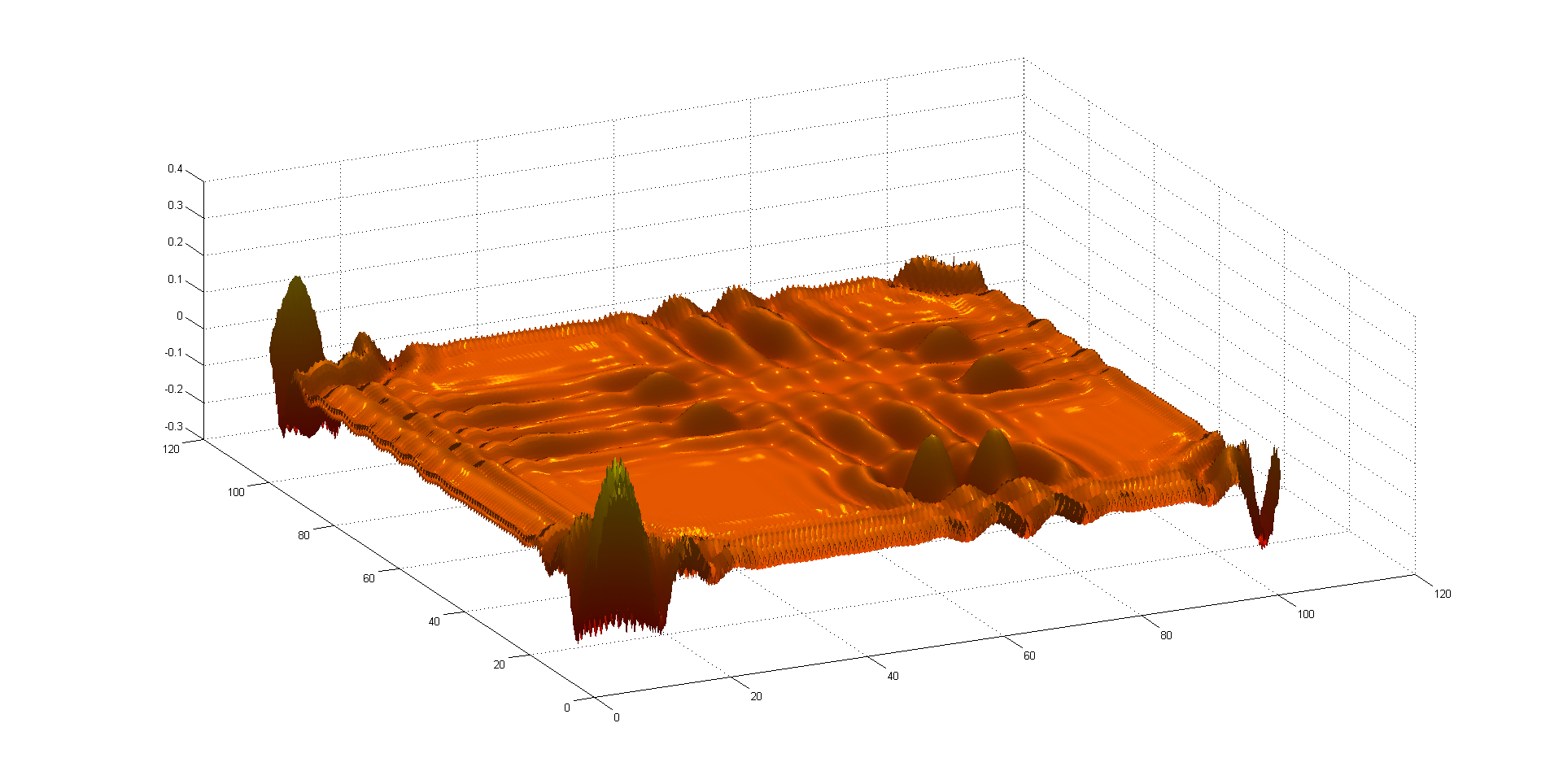}
    \caption{Solution to \eqref{eq:ourIBVP_2D}, \eqref{eq:IC3_2D} for $r=10$ at $t=20$}
    \label{fig:IC3_t=10}
\end{center}
\end{figure}

\begin{figure}
\begin{center}
    \includegraphics[keepaspectratio, width=8cm]{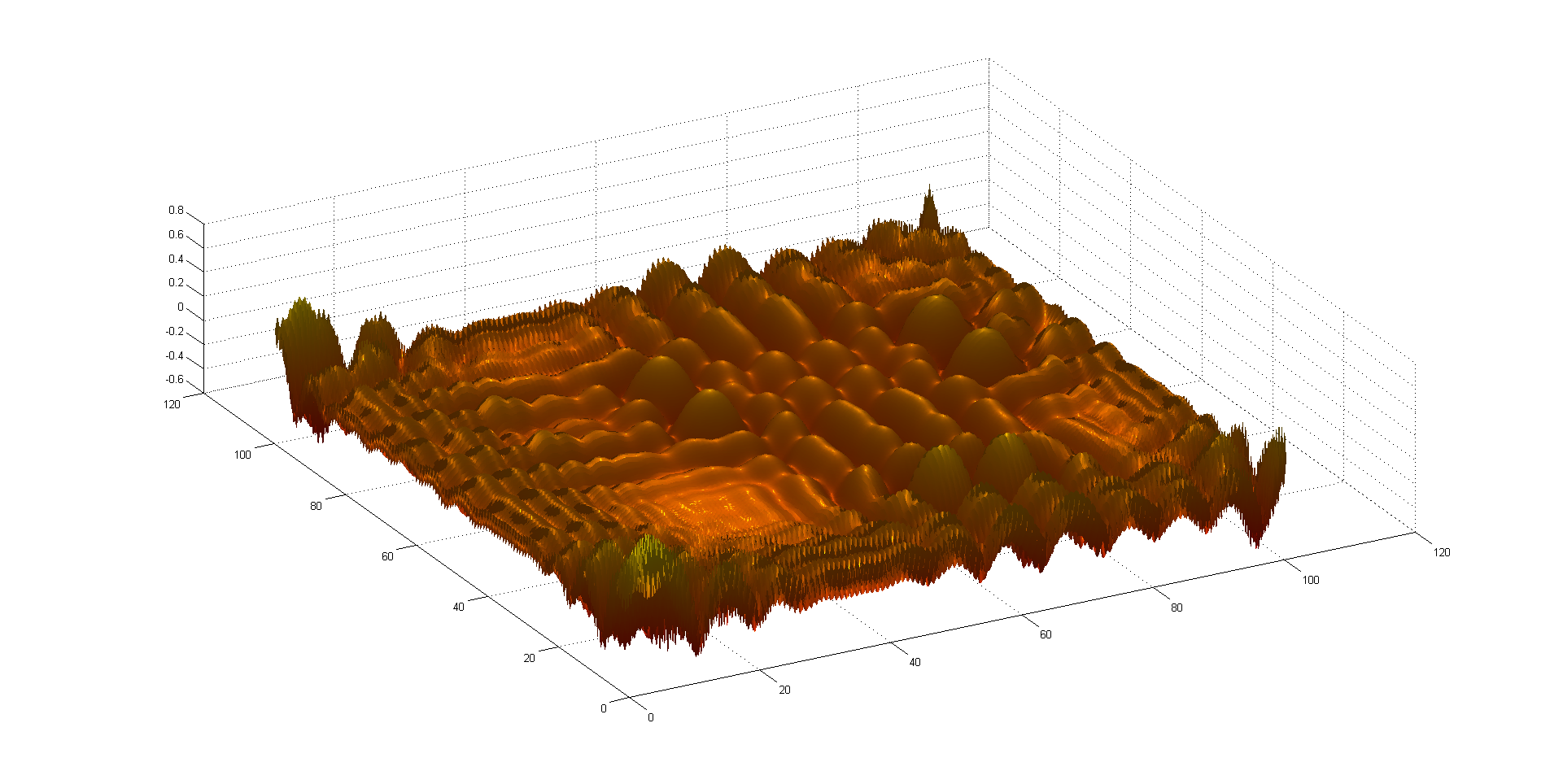}
    \caption{Solution to \eqref{eq:ourIBVP_2D}, \eqref{eq:IC3_2D} for $r=10$ at $t=30$}
    \label{fig:IC3_t=20}
\end{center}
\end{figure}

Figures \ref{fig:IC1_t=0}-\ref{fig:IC1_t=30} depict the solution
using the initial condition (\ref{eq:IC1_2D}), Figs.
\ref{fig:IC2_t=0}-\ref{fig:IC2_t=30} show it for (\ref{eq:IC2_2D})
and Figs. \ref{fig:IC3_t=0}-\ref{fig:IC3_t=20} represent it with
(\ref{eq:IC3_2D}). The numerical solutions are exhibited at discrete
time steps $t=0$, $t=20$ and $t=30$.

Our first impression might be the jaggedness of the graph of the
solution. It is so, because we did not use an aliasing method (like
zero padding or phase shift).  It is experienced from the graphs
that aliasing is much more prominent in case of the two-dimensional
problem. With the decrease of parameter $r$ in the nonlinear partial
differential equation \eqref{eq:ourIBVP_2D}, the simulation became
stable for longer time interval, similarly to the one-dimensional
case.

\section{Conclusion}

We have analyzed one- and two-dimensional evolution equations in the
context of amorphous thin film growth. It is found that the dynamics
depends on the initial wavelength and amplitude. Three different
behaviors are predicted. For large enough $\lambda(0)$ and $A(0)$,
the surface growth without limit, and in this case  $A(t)$ behaves
like $t$ and $\lambda(t)$ behaves like $\sqrt{t}$. If the initial
wavelength $\lambda(0)$ is small enough, the surface collapses at
finite time. This phenomena is a consequence of the presence of the
nonlinear conserved KPZ term. At some critical value of $\lambda $,
the surface structure does not change.

Some numerical solutions are presented for different value of
parameters $r$ and $A(0)$. For the same value of $r$, the surface
with bigger $A(0)$ starts to grow earlier than the surface with
smaller initial amplitude. For large $r$ the surface exhibits the
coarsening phenomena while for very small $r$  the surface shows
 chaotic phenomena. Note, that if $r\to 0$ then equation (\ref{eq:ourIBVP-1}) reduces to the
 Kuramoto-Sivashinsky equation, which exhibits spatiotemporal chaotic
 phenomena.
The numerical simulations for two-dimension in case of $r=10$
present the coarsening phenomena in agreement with the analytical
result. Further analytical investigation  will be done on the
interplay between the conserved KPZ term and nonconserved KPZ term
for general dimension $D$.
\section*{Acknowledgments}
This research was supported by the European Union and the Hungarian
State, co-financed by the European Regional Development Fund in the
framework of the GINOP-2.3.4-15-2016- 00004 project, aimed to
promote the cooperation between the higher education and the
industry. The authors acknowledge support by PHC-Balaton Number
34494UG, National Research, Development and Innovation Office within
the T\'ET\_14\_FR-1-2015-0004 project by 1.468 M Ft.


\section*{References}

\begin{thebibliography}{99}
\bibitem{V1}
J. Villain, J. Phys. I France 1, 19 (1991).
\bibitem{EH}
G. Ehrlich, F. G. Hudda, J. Chem. Phys. 44, 1039 (1966).
\bibitem{SS}
R. L. Schwoebel, E. J. Shipsey, J. Appl. Phys. 37,  3682 (1966).
\bibitem{KPS}
J. Krug, M. Plischke, M. Siegert, Phys. Rev. Lett. 70, 3271  (1993).
\bibitem{JOHG}
M. D. Johnson, C. Orme, A. W. Hunt, D. Graff, J. Sudijono, L. M.
Sander, Phys. Rev. Lett. 72, 116 (1994).
\bibitem{PMSKP}
O. Pierre-Louis, C. Misbah, Y. Saito, J. Krug, P. Politi, Phys. Rev.
Lett. 80, 4221 (1998).
\bibitem{uj}
F. L. Forgerini, R. Marchiori, Biomatter, 4:e28871; PMID: 24751679
(2014)
\bibitem{uj2}
J. Munoz-García, L. Vázquez, M. Castro, R. Gago, A. Redondo-Cubero,
A. Moreno-Barrado, R. Cuerno, Mater. Sci. Eng. R Rep. 86, 1 (2014)
\bibitem{ETB}
J. W. Evans, P. A. Thiel, M. C. Bartelt, Surface Science Reports 61,
1 (2006).
\bibitem{BMV}
 I. Bena,  C. Misbah,  A. Valance, Phys. Rev. B. 47, 7408 (1993).
  \bibitem{US}
 M. Uwaha,  M. Sato, Europhys. Lett. 32, 639 (1995).
 \bibitem{PGPM}
 S. Paulin, F. Gillet, O. Pierre-Louis, C. Misbah, Phys. Rev. Lett. 86, 5538 (2001).
 \bibitem{PM}
 P. Politi, C. Misbah, Phys. Rev. Lett. 92, 090601 (2004).


\bibitem{[6]}
T. Frisch, A. Verga, Phys. Rev. Lett. 96, 166104 (2006).

\bibitem{[10]}
 {F. Gillet, O. Pierre-Louis, C. Misbah},
 {Europ. Phys. J. B. 18}, 519 (2000).

\bibitem{[5]}
{F. Gillet, Z. Csahok, C. Misbah},  {Phys. Rev. B  63}, 241401
(2001).

\bibitem{[11]} {Z. Csahok,  C. Misbah, A. Valance},
  {Physica D}
  {128}, 87 (1999).

\bibitem{PobA}
  P. Politi, D. ben-Avraham, Physica D 238, 156 (2009).

  \bibitem{RLH}
M. Raible, S. J. Linz, P. H\"{a}nggi, Phys. Rev. E 62, 1691 (2000).

\bibitem{CCVG}
M. Castro, R. Cuerno, L. V\'{a}zquez, R. Gago, Phys. Rev. Lett. 94,
016102 (2005).
\bibitem{MCC}
J. Mu\~{n}osz-Garcia, R. Cuerno, M. Castro, Phys. Rev. E 74,
050103-1 (2006).

\bibitem{TLDS}
P. I. Tamborenea, Z.-W. Lai, S. Das Sarma, Surface Science 267, 1
(1994).

\bibitem{Gilding} B. H. Gilding, M. Guedda, R. Kersner, J. Math.
Anal. Appl. 284, 733 (2003).


 \bibitem{Cox2002}
 S. M. Cox, P. C. Matthews. Exponential Time Differencing for Stiff Systems.
Journal of Computational Physics, 176(2), 430 (2002).

\end{thebibliography}
\end{document}